\documentclass[preprint,showpacs,preprintnumbers,amsmath,amssymb]{revtex4}

\usepackage{graphicx}
\usepackage{dcolumn}
\usepackage{bm}

\begin{document}

\preprint{}

\title{Qubit-Qutrit 
Separability-Probability Ratios}

\author{Paul B. Slater}%
\email{slater@kitp.ucsb.edu}
\affiliation{%
ISBER, University of California, Santa Barbara, CA 93106\\
}%
\date{\today}

\begin{abstract}
Paralleling our recent computationally-intensive (quasi-Monte Carlo) 
work for the case $N=4$
(quant-ph/0308037),
we undertake the task for $N=6$ of computing to
high numerical accuracy, the formulas
of Sommers and \.Zyczkowski (quant-ph/0304041) for the $(N^2-1)$-dimensional volume and $(N^2-2)$-dimensional 
hyperarea of the (separable and nonseparable) 
$N \times N$ density
matrices, based on the Bures (minimal monotone) metric --- and 
also their
analogous 
formulas (quant-ph/0302197) for the (non-monotone) flat Hilbert-Schmidt
metric. With the same seven {\it billion} 
well-distributed 
(``low-discrepancy'') sample points,
we estimate the {\it unknown} volumes and hyperareas based on 
five additional (monotone) metrics of interest, including the Kubo-Mori
and Wigner-Yanase. Further, we 
estimate all of 
these seven volume and seven hyperarea 
(unknown) quantities when restricted
to the {\it separable} density matrices. The ratios of separable volumes 
(hyperareas) to separable {\it plus}
 nonseparable volumes (hyperareas) 
yield estimates of the {\it separability probabilities} of
generically rank-six (rank-five) density matrices.
The (rank-six) separability probabilities obtained based on the 
35-dimensional volumes 
appear to be --- {\it independently} of the metric
(each of the seven 
inducing Haar measure) employed --- {\it twice} as large as those 
(rank-five ones) 
based on the 34-dimensional
 hyperareas. (An additional  estimate --- 33.9982 --- of 
the ratio of the rank-6 Hilbert-Schmidt separability probability to the 
rank-{\it 4} one is quite clearly close to integral too.)  The doubling relationship also appears to hold 
S(4.15for the $N=4$ case for the Hilbert-Schmidt metric, but not the others. We 
fit simple {\it exact} formulas to
our estimates of the Hilbert-Schmidt
{\it separable} volumes and hyperareas in both the $N=4$ and $N=6$ cases.
\end{abstract}

\maketitle

\section{Introduction}
In Part I of the influential paper \cite{ZHSL,zycz2}, ``Volume of the set of mixed entangled
states'', \.Zyczkowski, Sanpera, Horodecki and Lewenstein considered 
the 
``question of how many entangled or, respectively, separable states are there in the set of all quantum states''. They cited philosophical, practical and physical reasons for doing so. They gave a qualitative argument 
\cite[sec. III.B]{ZHSL} --- contrary 
to their 
initial supposition --- that the measure of separable states could not be strictly zero. There has since been considerable work 
\cite{slatersilver,qubitqutrit,slaterA,slaterC,slaterSEP,firstqubitqubit,slaterprecursor,clifton,sepsize1,sepsize2,sepsize3,szarek}, using various forms of measures, 
 to determine/estimate the ``volume of separable states'', as well as the volume of separable {\it and} nonseparable states \cite{hans1,hilb2}, and hence probabilities of separability.  One somewhat surprising development has been the 
(principally 
 numerical) indication --- two independent estimates being 0.137884
\cite{slatersilver} and 0.138119 (sec.~\ref{N4qubitqubit} below) 
that the volume of separable states itself 
can take on a notably 
elegant form, in particular,  $(\sqrt{2}-1)/3 \approx 0.138071$, for the case of qubit-qubit pairs endowed with
the {\it statistical distinguishability} metric (four times the Bures metric).
(However, there seems to be a paucity of ideas on how to {\it formally}
prove or disprove such a conjecture.)
The research reported below was undertaken initially with the specific purpose of finding whether a putative 
comparably elegant formula for the volume of separable
qubit-qutrit pairs might exist. We will report below (sec.~\ref{wellfitting}) 
 the obtaining of certain possible formulas that fit our numerical results well, but
none of such striking simplicity (nor none that extends it, in any natural apparent fashion). But we also obtain some new-type
 results of substantial independent interest.

In a recent highly comprehensive 
analysis \cite{hans1} (cf. \cite{hansrecent}), Sommers and \.Zyczkowski obtained
``a fairly general expression for the Bures volume of the submanifold of the states of rank $N-n$ of the set of complex ($\beta=2$) or real ($\beta=1$)
$N \times N$ density matrices
\begin{equation} \label{HZ1}
S^{(\beta,Bures)}_{N,n}= 2^{-d_{n}} \frac{\pi^{(d_{n}+1)/2}}{\Gamma((d_n+1)/2)} \Pi^{N-n}_{j=1} \frac{\Gamma(j \beta/2) \Gamma[1+(2 n+j-1) \beta/2]}{\Gamma[(n+j) \beta/2] \Gamma[1+(n+j-1) \beta/2]},
\end{equation}
where $d_{n}= (N-n) [1+(N+n-1) \beta/2] -1$ represents the dimensionality of the
manifold \ldots for $n=0$ the last factor simply equals unity and (\ref{HZ1})
gives the Bures volume of the entire space of density matrices, equal to that
of a $d_{0}$-dimensional hyper-hemisphere with radius 1/2. In the case
$n=1$ we obtain the volume of the surface of this set, while for $n=N-1$ we get the volume of the set of pure states \ldots which for $\beta=1(2)$ gives correctly the volume of the real (complex) projective space of dimensions
$N-1$'' \cite{hans1}.
The Bures metric on various spaces of density matrices ($\rho$) 
has been widely
studied \cite{hubner1,hubner2,ditt1,ditt2}. In a broader context, it serves
as the {\it minimal} monotone metric \cite{petz1}.

In Part II of \cite{ZHSL,zycz2}, \.Zyczkowski put forth a certain
proposition. It was that ``the link between the purity of the mixed states
and the probability of entanglement is not sensitive to the measure [on the space of $N \times N$ density matrices] used''. His assertion was  based on
comparisons between a unitary product measure and an orthogonal product 
measure for the (qubit-qubit) case $N=4$ \cite[Fig. 2b]{zycz2}.
The {\it participation ratio} --- $1/\mbox{Tr}(\rho^{2})$ --- was used as the measure of purity.
\section{Separability-Probability Ratios}
In this study, we present (sec.~\ref{ratios}) 
numerical evidence --- limited largely to the 
specific (qubit-qutrit) case $N=6$ --- for a 
somewhat related 
proposition (which appears to be possibly 
topological in nature \cite{witten}).
It is that a certain ``ratio 
of ratios'' 
\begin{equation} \label{ratofrat}
\Omega^{metric} \equiv \frac{R^{metric}_{sep+nonsep}}{R^{metric}_{sep}}
\end{equation}
 is equal to 2, {\it independently} of the
measure used --- where the possible measures (including the 
just-discussed Bures) 
is comprised of 
{\it volume elements} (all incorporating {\it Haar} measure as a factor) of certain {\it 
metrics} defined on the $N \times N$ density
matrices.
Here by 
\begin{equation}
R^{metric}_{sep+nonsep} \equiv \frac{S^{(2,metric)}_{N,1}}{S^{(2,metric)}_{N,0}},
\end{equation}
is indicated
the ratio of the hyperarea of the $(N^2-2)$-dimensional boundary of the $(N^2-1)$-dimensional
convex set ($C_{N}$) of $N \times N$ density matrices
 to the total 
volume of $C_{N}$. Further, 
\begin{equation}
R^{metric}_{sep} \equiv \frac{\Sigma^{(2,metric)}_{N,1}}{\Sigma^{(2,metric)}_{N,0}}
\end{equation}
is the same type of 
hyperarea-volume ratio, 
but now restricted to the (classical/non-quantum) 
subset of $C_{N}$ composed of the {\it separable}
 states \cite{werner} 
(which we designate using $\Sigma$ rather than $S$).
A simple algebraic 
rearrangement of quotients then reveals that $\Omega^{metric}$ 
(\ref{ratofrat}) is also interpretable as
the 
ratio 
\begin{equation}
\Omega^{metric} \equiv \frac{P_{N}^{[metric,rank-N]}}{P_{N}^{[metric,rank-(N-1)]}}
\end{equation}
of the {\it separability probability} of the totality of 
(generically rank-$N$) 
states in $C_{N}$ 
\begin{equation}
P_{N}^{[metric,rank-N]} \equiv 
\frac{\Sigma^{(2,metric)}_{N,0}}{S^{(2,metric)}_{N,0}} 
\end{equation}
to the separability probability 
\begin{equation}
P_{N}^{[metric,rank-(N-1)]} \equiv
\frac{\Sigma^{(2,metric)}_{N,1}}{S^{(2,metric)}_{N,1}}
\end{equation}
of the 
(generically rank-$(N-1)$) states 
that lie  on the boundary of $C_{N}$.

\section{Metrics of interest}
Let us apply the \.Zyczkowski-Sommers Bures formula 
(\ref{HZ1}) to the two cases that will be 
of specific interest in this study,
$N=6$, $ n=0$, $\beta=2$ and $N=6$, $n=1$, $\beta=2$ --- that is, the Bures 35-dimensional volume and 
34-dimensional hyperarea of the {\it complex} 
 $6 \times 6$ density matrices. (It would, of course, also be of interest to
study the {\it real} case, $\beta=1$, though we have not undertaken any work
in that direction.)
We then have that 
\begin{equation} \label{m1}
S^{(2,Bures)}_{6,0}= \frac{{\pi }^{18}}{12221326970165372387328000} \approx 
7.27075 \cdot {10}^{-17}
\end{equation}
and
\begin{equation} \label{m2}
S^{(2,Bures)}_{6,1}= \frac{{\pi }^{17}}{138339065763438059520000} \approx 2.04457 \cdot {10}^{-15}.
\end{equation}
Here,  we are able
(somewhat paralleling our recent work for the qubit-qubit case $N=4$ 
\cite{slatersilver}, but in a 
rather
more systematic manner {\it ab initio} 
than there), through 
advanced numerical (quasi-Monte Carlo/quasi-random) methods, to reproduce 
both of these values (\ref{m1}), (\ref{m2}), to a considerable accuracy. 
At the same time, we compute numerical values --- it would seem reasonable
to presume, at least initially,
with roughly the same level of accuracy --- of these two quantities, but for the replacement of the Bures metric by five other {\it monotone} metrics of
interest. These are the Kubo-Mori \cite{hasegawa,petz3,michor,streater},
(arithmetic) average \cite{slatersilver}, Wigner-Yanase \cite{gi,wy,luo,luo2},
 Grosse-Krattenthaler-Slater  
(GKS) \cite{KS} (or ``quasi-Bures'' \cite{gillmassar})
 and (geometric) average monotone 
metrics --- the two ``averages'' being formed from the minimal 
(Bures) and {\it maximal} (Yuen-Lax \cite{yuenlax}) 
monotone metrics, following the suggested procedure in
\cite[eq. (20)]{petz2}. No proven 
formulas, such as (\ref{HZ1}), are presently available for these 
other various quantities, 
although our research in \cite{slatersilver} had suggested  that
the Kubo-Mori volume of the $N \times N$ density matrices is expressible as
\begin{equation} \label{KMgeneral}
S^{(2,KM)}_{N,0} = 2^{N(N-1)/2} S^{(2,Bures)}_{N,0},
\end{equation}
which for our case of $N=6$ would give
\begin{equation} \label{KM32768}
S^{(2,KM)}_{6,0} = 32768 S^{(2,Bures)}_{6,0}.
\end{equation}

In light of 
the considerable attention recently devoted to the (Riemannian, but 
 {\it non}-monotone \cite{ozawa}) 
Hilbert-Schmidt metric \cite{hansrecent,hilb1,hilb2},
including the availability of exact volume and hypersurface formulas 
\cite{hilb2},
we include it in supplementary analyses too.
Further, we estimate for all these seven (six monotone and one 
non-monotone) 
 metrics the (unknown) 35-dimensional volumes and
34-dimensional hyperareas restricted to the {\it separable} $2 \times 3$ and
$3 \times 2$ systems. Then, we can, obviously, by taking ratios 
of separable quantities to their separable {\it plus} nonseparable 
counterparts obtain ``probabilities of separability'' --- a topic which was first investigated in \cite{ZHSL}, 
and studied further, using the Bures metric, in \cite{slatersilver,slaterA,slaterC,qubitqutrit}.
\section{Two forms of partial transposition}
We will employ the convenient Peres-Horodecki necessary {\it and} 
sufficient positive 
partial
transposition criterion for separability \cite{asher,michal} --- asserting that a $4 \times 4$ or $6 \times 6$ density matrix is separable if and only if all the eigenvalues of its
partial transpose are nonnegative. (In the $4 \times 4$ [qubit-qubit] 
case, it simply suffices to test the determinant of the partial transpose 
for nonnegativity \cite{sanpera,verstraete}.) But in the $6 \times 6$ case, we have the qualitative difference that partial transposes can be determined in 
(at least) {\it 
two} inequivalent ways, either by transposing in place, in the natural
blockwise manner, 
the {\it nine} $2 \times 2$ submatrices or the {\it four} 
$3 \times 3$ submatrices \cite[eq. (20)]{michal}. (Obviously, such a nonuniqueness arises in a bipartite system only if the dimensions of the two parts are unequal.) 
We will throughout this study --- as in \cite{qubitqutrit} --- at the expense of added computation, 
analyze results using {\it both} forms of partial transpose.

It is our anticipation --- although yet 
without a formal demonstration  --- that in the limit of 
large sample size, the two sets of (separable volume and separable hyperarea) results 
of interest here should  converge
to true {\it common} values.
Now, the author must admit that he initially thought that it made no difference at all in which of the two ways the partial transpose was taken, that is,
a $6 \times 6$ density matrix would either pass or fail {\it both} tests. Also, this seems to be a common attitude in the quantum information community 
(as judged by a number of personal reactions [cf. \cite[fn. 2]{ZHSL}]).
Therefore, we present below a specific 
 example of a $6 \times 6$ density 
matrix ($\rho_{1}$) 
 that remains a density matrix if its four $3 \times 3$ blocks are transposed, but not its nine $2 \times 2$ blocks, since the latter 
result has a {\it 
negative}
eigenvalue (-0.00129836).
\begin{equation} \label{counterexample}
\begin{pmatrix}
\frac{2}{9}& 0 & 0 & 0 & 0 & 0 \\
   0 & \frac{1}{7} & 0 & 0 & 0 &
   -\frac{1}{24}   + \frac{\imath }{38} \\
  0 & 0 & \frac{1}{5} & \frac{\imath }{23} & \frac{-\imath }{41} &
   - \frac{1}{10}   - \frac{\imath }{21} \\
  0 & 0 & \frac{-\imath }{23} & \frac{1}{7} & 0 & 
   \frac{\imath }{13} \\
  0 & 0 & \frac{\imath }{41} & 0 & \frac{1}{6} & 0 \\
  0 & - \frac{1}{24}   - \frac{\imath }{38} &
   -\frac{1}{10}   + \frac{\imath }{21} &
   \frac{-\imath }{13} & 0 & \frac{79}{630} \\
\end{pmatrix}
\end{equation}
K. \.Zyczkowski has pointed out that the question of whether a given state
$\rho$ is entangled or not depends crucially upon the decomposition of the composite Hilbert space $H_{A} \otimes H_{B}$ (cf. \cite{zanardi,caban}). For instance, for the simplest 
$2 \times 2$ case, the maximally entangled Bell state becomes ``separable'', he points out, if one considers entanglement with respect to another division
of the space, {\it e. g.}  $A'=\{\Phi_+,\Phi_-\},B'=\{\Psi_+,\Psi_-\}$. So, it should not be surprising, at least in retrospect, 
 that some states are separable with respect to one form of partial transposition, and not the other.
In the course of examining this issue, we found that if one starts with
an arbitrary $6 \times 6$ matrix ($M$), and alternates the two forms of partial transpostion on it, after twelve ($=2 \times 6$) iterations of this process, one arrives back at the {\it original} $ 6 \times 6$ matrix. So, in group-theoretic terms, if we denote the three-by-three operation by $a_{3}$ and the two-by-two operation by  $a_{2}$, we have
idempotency, $a_{2}^2=a_{3}^2=I$ and $(a_{2} a_{3})^6 = (a_{3} a_{2})^6 =I$. Further, one can go from the
partial transpose  $a_{3}(M)$ to the partial transpose 
$a_{2}(M)$ {\it via} the matrix corresponding to the 
permutation $\{1,4,2,5,3,6\}$.

Further, we constructed the related density matrix ($\rho_{2}$)
\begin{equation}
\begin{pmatrix}
\frac{2}{9} & 0 & 0 & 0 & 0 & 0 \\
0 & \frac{1}{7} & 0 & 0 & \frac{\imath }{23} &
   \frac{-\imath }{41} \\ 
 0 & 0 & \frac{1}{5} & 0 &  -\frac{1}{24}   + 
    \frac{\imath }{38} & 0 \\ 
 0 &  0  &  0 &  \frac{1}{7} & -  \frac{1}{10}   - 
    \frac{\imath }{21} & \frac{-\imath }{13} \\
  0 & \frac{-\imath }{23} & 
   -\frac{1}{24} - \frac{\imath }{38} & 
   -\frac{1}{10}   + \frac{\imath }{21} &
   \frac{1}{6} & 0 \\
0 & \frac{\imath }{41} & 0 & 
   \frac{\imath }{13} & 0 & \frac{79}{630} \\.
\end{pmatrix}
\end{equation}
Now, if $\rho_{2}$ is partially transposed using its nine $2 \times 2$ blocks,
it gives the identical matrix  as when $\rho_{1}$ is partially 
transposed using four $3 \times3$ blocks. But the six eigenvalues of 
$\rho_{1}$, that is, $ \{ 0.322635, 
0.222222, 0.1721, 0.149677, 0.119158, 0.0142076 \}$ are {\it not} the same
as the six eigenvalues of $\rho_{2}$, that is, $ \{0.300489, 
0.222222, 0.204982, 0.168304, 0.0992763, 0.00472644 \} $. 
So, there can be no unitary transformation taking $\rho_{1}$ to $\rho_{2}$.
(The possibility that $\rho_{1}$ and $\rho_{2}$ might have
 the same total measure(s) attached to them, can not formally be ruled out, however.)
\section{Research Design}
Our main analysis will take the form of a quasi-Monte Carlo (Tezuka-Faure 
\cite{tezuka,giray1})  
numerical integration over the 35-dimensional hypercube ($[0,1]^{35}$) and a 
34-dimensional subhypercube of it. In doing so, we implement 
 a parameterization of the $6 \times 6$ density matrices in terms of {\it thirty} Euler angles (parameterizing $6 \times 6$ unitary matrices) and {\it five} hyperspherical angles
(parameterizing the {\it six} eigenvalues --- constrained to sum to 1) 
\cite{sudarshan,toddecg}. We hold a single one of the five hyperspherical angles fixed in the 34-dimensional analysis, so that one of the six eigenvalues is 
always zero --- and the density matrix is generically of rank five.  The parameters are linearly transformed so 
that they each lie in the unit interval [0,1] and, thus, collectively in the
unit hypercube. The computations consumed approximately five months using
six PowerMacs in parallel, each generating a different segment of the
Tezuka-Faure sequence.
\subsection{Silver mean ($\sqrt{2}-1$) conjectures for $N=4$}
We have previously pursued a similar numerical analysis in investigating the separable and nonseparable volumes and hyperareas of the $4 \times 4$ density matrices \cite{slatersilver}. 
Highly accurate results (as gauged in terms of {\it known} 
Bures quantities \cite{hans1}) --- based on two {\it billion} points of a Tezuka-Faure (``low discrepancy'') 
sequence lying in the 15-dimensional hypercube --- led us to advance several strikingly simple conjectures.
For example, it was indicated  that the Kubo-Mori volume of separable and nonseparable states was exactly $64 =2^6$ times the known 
Bures volume. (The exponent 
6 is  expressible --- in terms of our general conjecture (\ref{KMgeneral}), 
relating the Bures and Kubo-Mori volumes --- as $N(N-1)/2$, with 
$N=4$.)
Most prominently, though, it was conjectured 
that the statistical distinguishability 
(SD) volume of separable states 
is $\frac{\sigma_{Ag}}{3}$  and $10 \sigma_{Ag}$ in terms of 
(four times) the Kubo-Mori
metric. Here, $\sigma_{Ag} = \sqrt{2}-1 \approx 0.414214$ is the ``silver 
mean'' \cite{christos,spinadel,gumbs,kappraff} (cf. \cite{markowsky}). 
The SD metric is identically four times the Bures metric
\cite{caves}. (Consequently, the SD 15-dimensional volume of the $4 \times 4$ complex density matrices
is $2^{15}$ times that of the Bures volume --- given by 
formula (\ref{HZ1}) for $N=4,n=0,\beta=2$ --- thus equaling the volume
of a 15-dimensional hyper-hemisphere with radius 1, rather than $\frac{1}{2}$
 as in the Bures case itself \cite{hans1}.)

Unfortunately, there appears to be little in the way of 
indications in the literature, as to how one might {\it formally} prove or disprove these conjectures --- ``brute
force'' {\it symbolic} integration seeming to be well beyond present technical/conceptual capabilities (cf. \cite[sec. 5]{sudarshan}. (Certainly, Sommers and \.Zyczkowski \cite{hans1} did not directly
employ symbolic integration methodologies in deriving the Bures volume, hyperarea,...for $N$-level [separable {\it and}  nonseparable] systems, but rather, principally, 
used concepts of random matrix theory.)
One approach we have considered in this regard \cite{slaterSEP} 
 is to parameterize the 15-dimensional
convex set of bipartite qubit states in terms of the weights used in
the expansion of the state in some basis of sixteen  extreme
separable $4 \times 4$ density matrices (cf. \cite{rudiger1}). 
For a certain basis composed of $SU(4)$ generators \cite{kk,byrdkhaneja,gen}, 
the associated $15 \times 15$
Bures metric tensor \cite{ditt1} is {\it diagonal} in form 
(having all entries equal) 
 at the fully mixed state \cite[sec.~II.F]{slaterSEP}. 
(Also, we have speculated that perhaps there is some way of ``bypassing''
the formidable computation of the Bures metric tensor, and yet being
able to arrive at the required volume element.) Perhaps, though, 
at least in the Bures/{\it minimal} 
monotone case, a proof might be based on the concept of ``minimal volume'' \cite{bayard,bowditch,bambah}.
\subsection{Formulas for monotone metrics}
The monotone metrics (of which we study five, in addition to the
Bures) can all be expressed
in the general form
\begin{equation}
g_{\rho}(X',X) 
 = \frac{1}{4} \Sigma_{\alpha,\beta} |\langle \alpha |X| \beta \rangle |^2 c_{monotone}(\lambda_{\alpha},\lambda_{\beta})
\end{equation}
(cf. \cite{hubner1,hubner2}).
Here $X,X'$ lie in the tangent space of all Hermitian $N \times N$ density
matrices $\rho$ and $|\alpha \rangle, \alpha =1, 2 \ldots$ are 
the eigenvectors
of $\rho$ with eigenvalues $\lambda_{\alpha}$.
Now, $c_{monotone}(\lambda_{\alpha},\lambda_{\beta})$ represents the specific {\it Morozova-Chentsov} function for the monotone metric in question \cite{petz2}. This function takes the form
for: (1) the Bures metric,
\begin{equation} \label{Bures}
c_{Bures}(\lambda_{\alpha},\lambda_{\beta}) = \frac{2}{\lambda_{\alpha} +\lambda_{\beta}};
\end{equation}
(2) the Kubo-Mori metric (which, up up to a scale factor, is 
the unique monotone
Riemannian metric with respect to which the {\it exponential} and {\it mixture}
connections are dual \cite{streater}),
\begin{equation} \label{KM}
c_{KM}(\lambda_{\alpha},\lambda_{\beta}) =\frac{\log{\lambda_{\alpha}}-\log{\lambda_{\beta}}}{\lambda_{\alpha}-\lambda_{\beta}};
\end{equation}
(3) the (arithmetic) average metric (first discussed in \cite{slatersilver}),
\begin{equation}
c_{arith}(\lambda_{\alpha},\lambda_{\beta}) = \frac{4 (\lambda_{\alpha}+\lambda_{\beta})}{\lambda_{\alpha}^2 + 6 \lambda_{\alpha} \lambda_{\beta} +\lambda_{\beta}^2};
\end{equation}
(4) the Wigner-Yanase metric (which corresponds
to a space of {\it constant curvature} \cite{gi});
\begin{equation}
c_{WY}(\lambda_{\alpha},\lambda_{\beta}) =\frac{4}{(\sqrt{\lambda_{\alpha}} +\sqrt{\lambda_{\beta}})^2};
\end{equation}
(5) the GKS/quasi-Bures metric (which yields the asymptotic redundancy for
universal quantum data compression \cite{KS});
\begin{equation}
c_{GKS}(\lambda_{\alpha},\lambda_{\beta})= \frac{{(\frac{\lambda_{\alpha}}{\lambda_{\beta}})}^{\lambda_{\alpha}/(\lambda_{\beta}-\lambda_{\alpha})}}{\lambda_{\beta}} e;
\end{equation}
and (6)  the (geometric) average metric (apparently previously unanalyzed),
\begin{equation} \label{geom}
c_{geom}(\lambda_{\alpha},\lambda_{\beta}) =\frac{1}{\sqrt{ \lambda_{\alpha} \lambda_{\beta}}}.
\end{equation}
(The results obtained below for the geometric average monotone metric
seem, in retrospect,
 to be of little interest, other than indicating --- that like the
maximal monotone (Yuen-Lax) metric itself \cite{slatersilver} --- volumes 
and hyperareas appear to be simply
{\it infinite} in magnitude.)
\section{Analyses}
\subsection{Volumes and Hyperareas Based on Certain Monotone Metrics} \label{secMonotone}
\begin{figure}
\includegraphics{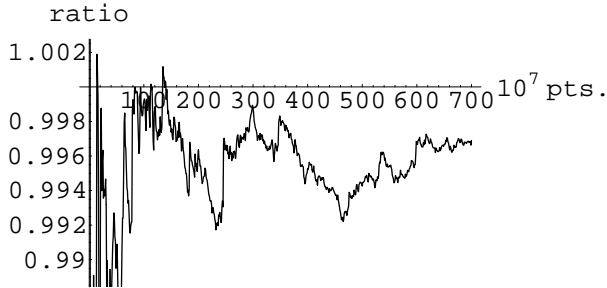}
\caption{\label{fig:graphBuresTotVol}Ratios of the cumulative estimates of
the 35-dimensional volume $S^{(2,Bures)}_{6,0}$ to its {\it known} value (\ref{m1}). For each addtional point shown --- as 
in all the subsequent plots --- {\it 
ten million} ($10^7$) values of the particular seven-billion-point 
Tezuka-Faure sequence have been generated.}
\end{figure}

\begin{figure}
\includegraphics{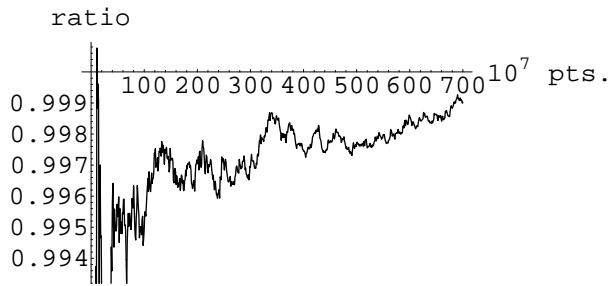}
\caption{\label{fig:graphBuresTotArea}Ratios of the cumulative estimates of
the 34-dimensional hyperarea 
$S^{(2,Bures)}_{6,1}$ to its known value (\ref{m2})}
\end{figure}

\begin{figure}
\includegraphics{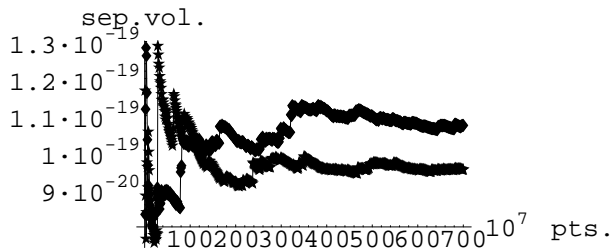}
\caption{\label{fig:graphBuressep}A {\it pair} of cumulative estimates of the 35-dimensional 
Bures 
volume of {\it separable}
qubit-qutrit states based on the {\it two} distinct
forms of partial transposition}
\end{figure}

\begin{figure}
\includegraphics{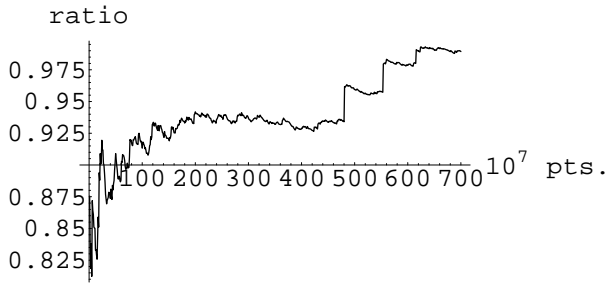}
\caption{\label{fig:graphKMBuresratio}Ratios of the cumulative estimates of
the 35-dimensional volume $S^{(2,KM)}_{6,0}$
 to its conjectured value of 
$32768   S^{(2,Bures)}_{6,0}$}
\end{figure}

Using 
the first seven  billion points of a Tezuka-Faure sequence, we
obtained the results reported in Tables~\ref{tab:table1}-\ref{tab:isoperimetric2} and Figs.~\ref{fig:graphBuresTotVol}-\ref{fig:ratioofratiosKM}. 
We followed the Bures 
formulas in \cite[secs. III.C, III.D]{hans1}, substituting for
(\ref{Bures}) the Morozova-Chentsov functions
given above (\ref{KM})-(\ref{geom}) to obtain the non-Bures  counterparts.

In Fig.~\ref{fig:graphBuresTotVol} 
we show the ratios of the cumulative estimates 
of the 35-dimensional volume 
$S^{(2,Bures)}_{6,0}$ to its {\it known} value (\ref{m1}). Each successive 
point is based on 
ten million ($10^7$) more systematically 
sampled values in the 35-dimensional hypercube than the 
previous point in the computational sequence.
In Fig.~\ref{fig:graphBuresTotArea}
we show the ratios of the cumulative estimates
of the 34-dimensional hyperarea 
$S^{(2,Bures)}_{6,1}$ to its known value (\ref{m2}). Each successive
point is based on
ten million more sampled values in the 34-dimensional hypercube than the 
previous point in the computational sequence.  The 
single Tezuka-Faure sequence we 
employ for all our purposes, however,  is specifically designed
as a {\it 35}-dimensional one --- of which we take an essentially arbitrary
34-dimensional {\it projection}. This is arguably 
a suboptimal strategy 
for generating well-distributed points in the 34-dimensional hypercube 
 (cf. \cite[sec. 7]{morokoff}), 
but it is certainly highly computationally convenient for us (since we avoid having to generate a totally 
new 34-dimensional sequence --- which would, we believe, 
 increase our computation time
roughly 50 percent), and seems to perform rather well.
(In fact, as discussed below, the {\it bias} of our estimates seems to be
--- contrary to expectations --- markedly less for the known [Bures and
Hilbert-Schmidt] 34-dimensional hyperareas than for the 
35-dimensional volumes.)

We also present a  joint plot (Fig.~\ref{fig:graphBuressep}) of the 
two sets of cumulative estimates of the Bures 
 volume of {\it separable} qubit-qutrit 
states based on {\it both} forms of partial transposition.
The estimates obtained using the four blocks of $3 \times 3$ submatrices,
in general, dominate those using nine blocks of $2 \times 2$ submatrices.

In Table~\ref{tab:table1}, we scale the estimates (which we denote using
$\tilde{S}$) 
of the volumes and hyperareas by the {\it known} values (\ref{m1}), (\ref{m2}) of
$S^{(2,Bures)}_{6,0}$ and $S^{(2,Bures)}_{6,1}$, while in Table~\ref{tab:table2} we scale these estimates by the {\it estimated} values ($7.22904 \cdot 10^{-17}$  and $2.03991 \cdot 10^{-15}$) of these two quantities.
(We use {\it both} approaches because we are uncertain as to which may be more revealing as to possible exact formulas --- an approach 
suggested by our work in \cite{slatersilver}.) The 
results for the geometric average monotone metric in Table~\ref{tab:table1} 
appear to be divergent. 
We might speculate that the 
middle four scaled hyperareas
in the last column of Table~\ref{tab:table1}
correspond to the actual values $7 \cdot 13/2=45.5, 
2^2 \cdot 5^2/3 \approx 31.333, 3 \cdot 13/4=9.75$ and $7/2=3.5$, and for the
second column that we have $132 = 12 \cdot 11$ and 12, as actual values.

\begin{table}
\caption{\label{tab:table1}Scaled estimates based on the Tezuka-Faure sequence of 7 billion points of 
 the 35-dimensional volumes and 34-dimensional
hyperareas of the $6 \times 6$ density matrices, using several monotone metrics. The scaling factors are the {\it known}
values of the volume and 
hyperarea for the Bures metric, given by (\ref{HZ1}), and more specifically
for the cases: $N=6$;  $n=0,1$; and $\beta=2$ by
(\ref{m1}) and (\ref{m2}).}
\begin{ruledtabular}
\begin{tabular}{rrr}
metric & $\tilde{S}_{6,0}^{(2,metric)}/S_{6,0}^{(2,Bures)}$ & 
$\tilde{S}_{6,1}^{(2,metric)}/S_{6,1}^{(2,Bures)}$ \\
\hline
Bures & 0.996899  & 0.999022 \\
KM & 32419.4  & 45.4577  \\
arith & 621.714  & 31.291  \\
WY & 131.711  & 9.76835  \\
GKS & 12.4001  & 3.55929 \\
geom & $2.80011   \cdot 10^{44}$ & $1.44011   \cdot 10^{14}$  \\
\end{tabular}
\end{ruledtabular}
\end{table}
\begin{table}
\caption{\label{tab:table2}Scaled estimates based on the Tezuka-Faure sequence of 7 billion points of the 35-dimensional volumes and 34-dimensional
hyperareas of the $6 \times 6$ density matrices, 
using several monotone metrics. The scaling factors  are the {\it estimated}
values ($\tilde{S}^{(2,Bures)}_{6,0} = 7.2482 \cdot 10^{-17}$  and 
$\tilde{S}^{(2,Bures)}_{6,1} = 2.04257   \cdot 10^{-15}$ ) of the volume and hyperarea for the Bures metric.}
\begin{ruledtabular}
\begin{tabular}{rrr}
metric & $\tilde{S}_{6,0}^{(2,metric)}/\tilde{S}_{6,0}^{(2,Bures)}$ & 
$\tilde{S}_{6,1}^{(2,metric)}/\tilde{S}_{6,1}^{(2,Bures)}$ \\
\hline
KM & 32520.3  & 45.5022 \\
arith & 623.648  & 31.3216 \\
WY & 132.121  & 9.77791  \\
GKS & 12.4387  & 3.56278   \\
geom & $2.80882  \cdot 10^{44}$ & $1.44152  \cdot 10^{14}$ \\
\end{tabular}
\end{ruledtabular}
\end{table}
In Tables~\ref{tab:table3} and Tables~\ref{tab:table4}, we report our estimates (scaled by the values obtained for the Bures metric) 
of the volumes and hyperareas of the $6 \times 6$ separable complex density matrices.
Let us note, however, that 
to compute the hyperarea of the {\it complete} boundary of the separable states, one must also include those $6 \times 6$ density matrices of {\it full}
 rank,
the partial transposes of which have a zero eigenvalue, with all other eigenvalues  being nonnegative 
 \cite{shidu}. (We do not compute this additional 
contribution here --- as we undertook to do in our lower-dimensional 
analysis \cite{slatersilver} --- as 
it would slow quite considerably the overall process in which we are engaged, since high-degree polynomials
would need to be 
solved at each iteration.)

 In 
\cite{slatersilver}, we had been led to conjecture that
that part of the 14-dimensional boundary of separable $4 \times 4$ density matrices consisting
generically of rank-{\it four} density matrices had 
SD hyperarea $\frac{55 \sigma_{Ag}}{39}$
and that part composed of rank-{\it three} density matrices, $\frac{43 \sigma_{Ag}}{39}$, for a 
total 14-dimensional boundary 
SD hyperarea of $\frac{98 \sigma_{Ag}}{39}$. We, then, sought to apply the
``Levy-Gromov isoperimetric inequality'' \cite{gromov} to the relation between the 
known and estimated SD
volumes and hyperareas of the separable and separable plus nonseparable
states \cite[sec. VII.C]{slatersilver}.

Restricting ourselves now to considering only the separable density matrices, 
for Table~\ref{tab:table3} we computed the partial transposes of the $6 \times 6$ density matrices by transposing in place the {\it four}  $3 \times 3$ submatrices, while
in Table~\ref{tab:table4} we transposed in place the nine 
 $2 \times 2$ submatrices.
\begin{table}
\caption{\label{tab:table3}Scaled estimates based on the Tezuka-Faure sequence of 7 billion points of the 35-dimensional volumes and 34-dimensional
hyperareas of the {\it separable} $6 \times 6$ density matrices, using several monotone metrics. The scaling factors are the {\it estimated}
values ($\tilde{\Sigma}^{(2,Bures)}_{6,0} = 1.0739  \cdot 10^{-19}$ and $
\tilde{\Sigma}^{(2,Bures)}_{6,1}= 1.53932   \cdot 10^{-18}$) --- the true values being unknown --- of the separable 
volume  and hyperarea  for the Bures metric. To implement the Peres-Horodecki positive partial transposition criterion,
 we compute
the partial transposes 
of the four $3 \times 3$ submatrices (blocks) of the density matrix.}
\begin{ruledtabular}
\begin{tabular}{rrr}
metric & Bures-scaled separable volume  &  Bures-scaled separable (rank-5) hyperarea  \\
\hline
KM & 8694.79  & 9.43481  \\
arith & 220.75   & 10.6415  \\
WY & 55.3839  & 4.13924  \\
GKS & 7.97798   & 2.28649   \\
geom & $3.33872   \cdot 10^{32} $ & $3.61411  \cdot 10^{8}$\\
\end{tabular}
\end{ruledtabular}
\end{table}
\begin{table}
\caption{\label{tab:table4}Scaled estimates based on the Tezuka-Faure sequence of 7 billion points of the 35-dimensional volumes and 34-dimensional
hyperareas of the {\it separable} $6 \times 6$ density matrices, using several monotone metrics. The scaling factors are the {\it estimated}
values ($\tilde{\Sigma}^{(2,Bures)}_{6,0} = 9.54508   \cdot 10^{-20}$  and $\tilde{\Sigma}^{(2,Bures)}_{6,1}= 1.40208   \cdot 10^{-18}$) --- the true values being unknown --- of the separable volume and hyperarea 
for the Bures metric. To implement the Peres-Horodecki positive partial transposition criterion, we compute
the partial transposes of the {\it nine}  $2 \times 2$ submatrices (blocks)  of the density 
matrix.}
\begin{ruledtabular}
\begin{tabular}{rrr}
metric & Bures-scaled separable volume &  Bures-scaled separable (rank-5) hyperarea  \\
\hline
KM & 6465.86  & 9.0409 \\
arith & 218.602   & 10.3248   \\
WY & 55.5199 & 4.05201   \\
GKS & 7.92729  & 2.26136  \\
geom & $5.4299  \cdot 10^{35}$ & $4.35667  \cdot 10^{9}$ \\
\end{tabular}
\end{ruledtabular}
\end{table} 
In Table~\ref{tab:table5}, we only require the density matrix 
in question to pass {\it either} of the two tests, while 
in Table~\ref{tab:table6}, we require it 
to pass {\it both} tests for separability.
(Of the 7 billion points of the Tezuka-Faure 35-dimensional sequence so 
far generated, 
approximately 2.91 percent yielded density matrices passing the test for Table I, 2.84 percent  for
Table II, 4 percent  for Table III and 1.75 percent  for Table IV. K. 
\.Zyczkowski commented that ``it is not reasonable to ask about the probability that 
{\it both} partial transpositions are simultaneously positive, since one should not mix two different physical problems together''.)
\begin{table}
\caption{\label{tab:table5}Scaled estimates based on the Tezuka-Faure sequence of 7 billion points of the 35-dimensional volumes and 34-dimensional
hyperareas of the {\it separable} $6 \times 6$ density matrices, using several monotone metrics. The scaling factors are the {\it estimated}
values ($1.99772  \cdot 10^{-19}$  and $2.90956   \cdot 10^{-18}$)--- the true values being unknown --- for the Bures metric. 
A density matrix is included here if it passes {\it either} form of the positive partial transposition test.}
\begin{ruledtabular}
\begin{tabular}{rrr}
metric & Bures-scaled separable volume  & Bures-scaled separable (rank-5) 
hyperarea  \\
\hline
KM & 7735.7  & 9.30446  \\
arith & 221.689  & 10.5467   \\
WY & 55.8928  & 4.11453  \\
GKS & 7.99075  & 2.28089   \\
geom & $2.59619   \cdot 10^{35}$ & $2.29051  \cdot 10^{9}$  \\
\end{tabular}
\end{ruledtabular}
\end{table}
\begin{table}
\caption{\label{tab:table6}Scaled estimates based on the Tezuka-Faure sequence of 7 billion points of the 35-dimensional volumes and 34-dimensional
hyperareas of the {\it separable} $6 \times 6$ density matrices, using several monotone metrics. The scaling factors are the {\it estimated}
values ($3.06807  \cdot 10^{-21}$  and $3.78991  \cdot 10^{-32}$) --- the true values being unknown --- for the Bures metric. 
A density matrix is included here {\it only} 
if it passes 
{\it both} forms of the positive partial transposition test.}
\begin{ruledtabular}
\begin{tabular}{rrr}
metric & Bures-scaled separable volume  &  Bures-scaled separable (rank-5) hyperarea  \\
\hline
KM & 1800.19  & 3.99932 \\
arith & 92.7744   & 5.3548   \\
WY & 26.4785  & 2.55592  \\
GKS & 5.56969   & 1.77049   \\
geom & $3.96779   \cdot 10^{27}$ & $1.09937  \cdot 10^{7}$  \\
\end{tabular}
\end{ruledtabular}
\end{table}

In Table VII, we ``pool'' (average) the results for the separable volumes
and hyperareas reported in Tables III and IV, based on the two 
distinct forms of
partial transposition, to obtain possibly superior estimates of these
quantities, which presumably are actually one and the same {\it independent}
of the particular form of partial transposition.
\begin{table}
\caption{\label{tab:table7}Scaled estimates 
obtained by pooling the results from Tables III and
IV --- based on the two forms of partial transposition --- for 
the separable volumes and hyperareas. The Bures scaling factors
(pooled volume and hyperarea) are $ \tilde{\Sigma}^{(2,Bures)}_{6,0} = 1.0142   \cdot 10^{-19}$ 
and $ \tilde{\Sigma}^{(2,Bures)}_{6,1} = 1.4707  \cdot 10^{-18}$.}
\begin{ruledtabular}
\begin{tabular}{rrr}
metric & Bures-scaled separable volume & Bures-scaled separable (rank-5) hyperarea \\
\hline
KM & 7645.92   & 9.24704  \\
arith & 219.739  & 10.4905   \\
WY & 55.4479  & 4.09766   \\
GKS & 7.95413    & 2.27537  \\
geom & $2.55692   \cdot 10^{35}$ & $2.26584  \cdot 10^{9}$ \\
\end{tabular}
\end{ruledtabular}
\end{table}

In Fig.~\ref{fig:graphKMBuresratio}
we show the ratios 
of $\tilde{S}^{(2,KM)}_{6,0}$ 
to its conjectured value (\ref{KM32768}) 
of $32768  S^{(2,Bures)}_{6,0}$.

\subsection{Volumes and Hyperareas based on the Hilbert-Schmidt Metric} \label{secHS}

Along with the computations based on six distinct monotone metrics, reported
above in sec.~\ref{secMonotone}, 
we have at the same time carried out fully parallel analyses of the
(Riemannian, but 
non-monotone) Hilbert-Schmidt metric \cite{ozawa}. These have
 only been conducted
{\it after} an earlier less-extensive 
form of this analysis \cite{slaterprecursor},
reporting initial numerical estimates for the same six monotone 
metrics based on 600 million points of a 
Tezuka-Faure sequence,  was 
posted. At that stage of our research, we had --- with certainly 
some
regrets --- decided to {\it fully} redo the
computations reported there. This was done 
to avoid a (somewhat inadvertent) programming limitation (which seemed of minor importance at the time) --- a 
consequence essentially only 
of our, in time, 
having  understood how to greatly speed up the 
computations  --- of not being able to
sample {\it more} than two billion Tezuka-Faure points.
This fresh beginning (incorporating a much larger limitation, 
of which we here take advantage) 
allowed us then, as well,  to additionally fully 
include the Hilbert-Schmidt metric. (It is somewhat 
unfortunate, however, at this
point, that we had not conducted analyses based on the HS metric
for the $N=4$ qubit-qubit case, having restricted our earlier attention to monotone metrics only \cite{slatersilver} [cf. sec.~\ref{N4qubitqubit}].)

Prior to  Sommers and \.Zyczkowski reporting their exact formula
(\ref{HZ1})
for the Bures volume of the submanifold of the states
 of rank $N-n$ of the set of complex ($\beta=2$) or real ($\beta=1$)
$N \times N$ density matrices, they had obtained  fully analogous
formulas for the Hilbert-Schmidt metric, which for the specific 
volume ($n=0$) case gives 
\cite[eq. (4.5)]{hilb2},
\begin{equation}
S^{(2,HS)}_{N,0} = \sqrt{N} (2 \pi)^{N(N-1)/2} \frac{\Gamma(1) \ldots 
 \Gamma(N)}{\Gamma(N^2)}
\end{equation}
and hyperarea ($n=1$) case \cite[eq. (5.2)]{hilb2} gives,
\begin{equation}
S^{(2,HS)}_{6,1} = 
\sqrt{N-1} (2 \pi)^{N(N-1)/2} \frac{\Gamma(1) \ldots \Gamma(N+1)}{\Gamma(N) \Gamma(N^2-1)}.
\end{equation}

For the (qubit-qutrit) 
case $N=6$ under study in this paper, these give us, for the 
35-dimensional HS volume,
\begin{equation} \label{HSvol1}
S^{(2,HS)}_{6,0} =
\frac{{\pi }^{15}}
  {1520749664069126407256340000000\,{\sqrt{6}}} \approx 7.69334\,{10}^{-24}
\end{equation}
and for the 34-dimensional HS hyperarea
\begin{equation} \label{HShyperarea1}
S^{(2,HS)}_{6,1} = \frac{{\pi }^{15}}
  {8689998080395008041464800000\,{\sqrt{5}}} \approx 1.47483\,{10}^{-21}.
\end{equation}
So, as above, using the Bures metric, 
we can further gauge the accuracy of the Tezuka-Faure numerical
integration in terms of these {\it known} volumes and hyperareas.
(This somewhat alleviates the shortcoming of the Tezuka-Faure procedure 
in not lending itself to statistical testing in any straightforward manner.)

The estimated probability of separability is {\it greater} for the HS-metric
than for any monotone one. (The minimal monotone or Bures metric appears
to give the greatest probability in the 
nondenumerable class of monotone metrics. Also, the {\it maximal}
monotone metric seems to give a {\it zero} probability \cite{slatersilver}.)
Therefore, one might surmise that the much-discussed 
estimates of the sizes of the separable
neighborhoods \cite{sepsize1,sepsize2,sepsize3} --- which usually appear to be based on the 
HS or Frobenius metric --- surrounding 
the fully mixed state are on the rather generous side (cf. \cite{szarek}), 
relatively speaking.

\begin{figure}
\includegraphics{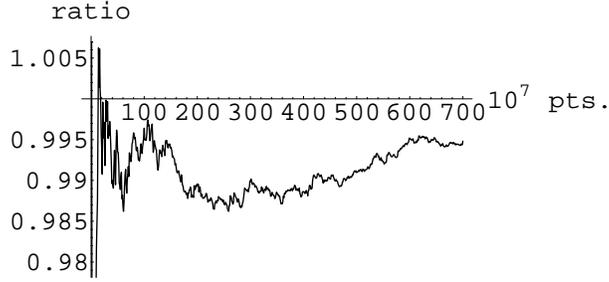}
\caption{\label{fig:graphHSTotVol}Ratios of the cumulative estimates of
the 35-dimensional Hilbert-Schmidt (Euclidean) volume 
$S^{(2,HS)}_{6,0}$ to its {\it known} value (\ref{HSvol1})}
\end{figure}

\begin{figure}
\includegraphics{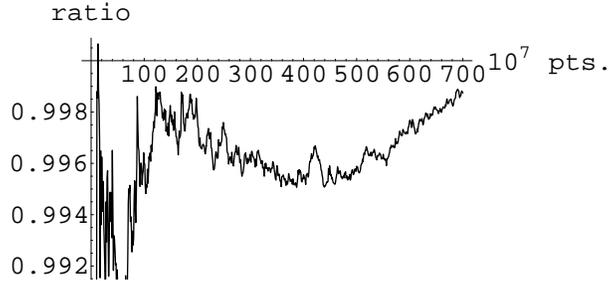}
\caption{\label{fig:graphHSTotArea}Ratios of the cumulative estimates of
the 34-dimensional Hilbert-Schmidt hyperarea 
$S^{(2,HS)}_{6,1}$ to its {\it known} value (\ref{HShyperarea1})}
\end{figure}

In Fig.~\ref{fig:graphHSTotVol}
we show --- paralleling Fig.~\ref{fig:graphBuresTotVol} --- the 
ratios of our  cumulative estimates
of $S^{(2,HS)}_{6,0}$ to its known value (\ref{HSvol1}).
In Fig.~\ref{fig:graphHSTotArea}
we show --- paralleling Fig.~\ref{fig:graphBuresTotArea} --- the 
ratios of the cumulative estimates
of $S^{(2,HS)}_{6,1}$ to its known value (\ref{HShyperarea1}).

\begin{figure}
\includegraphics{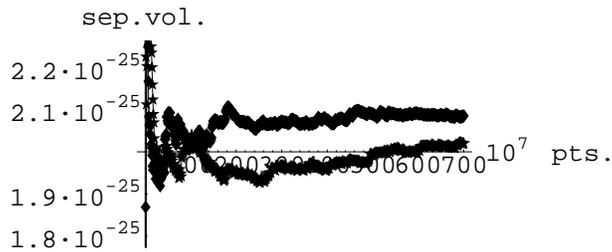}
\caption{\label{fig:graphHSsep}Cumulative estimates of the 
35-dimensional Hilbert-Schmidt volume of {\it separable}
qubit-qutrit states (for the two 
possible forms of partial transposition)}
\end{figure}

A plot (Fig.~\ref{fig:graphHSsep}) of the cumulative estimates of the Hilbert-Schmidt volume of separable qubit-qutrit states (for 
the two forms of partial transposition) is also presented.
(The ratio of the two cumulative estimates at the final [seven billion]
mark is 1.03236, while the comparable ratio is 1.12508 in the analogous
Bures plot [Fig~\ref{fig:graphBuressep}].)

In their two studies \cite{hans1,hilb2}, deriving exact formulas for the Bures and Hilbert-Schmidt volumes and hyperareas of the $N \times N$ density matrices, Sommers and \.Zyczkowski also 
explicitly derived expressions for the {\it ratios} of
$(N^2-2)$-dimensional hyperareas to $(N^2-1)$-dimensional volumes. 
These were \cite[eq. (4.20)]{hans1}
\begin{equation} \label{gammaBures}
\gamma_{Bures,N}= \frac{S^{(2,Bures)}_{N,1}}{S^{(2,Bures)}_{N,0}} 
= \frac{2}{\sqrt{\pi}} \frac{\Gamma (N^2/2)}{\Gamma(N^2/2-1/2)} N
\end{equation}
and \cite[eq. (6.5)]{hilb2}
\begin{equation} \label{gammaHS}
\gamma_{HS,N}= \frac{S^{(2,HS)}_{N,1}}{S^{(2,HS)}_{N,0}}
= \sqrt{N(N-1)} (N^2-1).
\end{equation}
In the $N=6$ Bures case, this ratio (equivalently what we have earlier denoted
$R^{Bures}_{sep+nonsep}$) is
\begin{equation}
\gamma_{Bures,6} \equiv \frac{S^{(2,Bures)}_{6,1}}{S^{(2,Bures)}_{6,0}} = \frac{2^{34}}{3^2 \cdot 5 \cdot 11 \cdot 19 \cdot 23 \cdot 29 \cdot 31  \pi}
= \frac{17179869184}{194467185 \pi} \approx 28.1205,
\end{equation}
and in the Hilbert-Schmidt case (equivalently $R^{HS}_{sep+nonsep}$),
\begin{equation}
\gamma_{HS,6} \equiv \frac{S^{(2,HS)}_{6,1}}{S^{(2,HS)}_{6,0}} 
= 35 \sqrt{30} \approx 191.703.
\end{equation}
(The Bures ratio grows 
proportionally with the dimensionality ($D=N^2-1$) of the 
$N \times N$ density matrices as $D$ (for large $N$)
\cite[sec.~IV.C]{hans1}, and as $D^{3/2}$
for the Hilbert-Schmidt ratio \cite[sec.~VI]{hilb2}.)
Our sample estimates for these two quantities are 28.1804 and 192.468,
respectively.
In Table~\ref{tab:isoperimetric}, we report these estimates, as well as 
the  sample estimates for the other
five  metrics under study here. We also list the two known values
and also give the corresponding ratios of hyperarea to volume for
a 35-dimensional {\it Euclidean} 
ball having: (1) the same volume as for 
the metric in question;
 and (2) the same hyperarea.
{\it Only} for the (flat) 
HS-metric are these last two ratios {\it less} 
than unity (cf. \cite[sec. VI]{hilb2}).
\begin{table}
\caption{\label{tab:isoperimetric}Sample estimates of the ratio ($R^{metric}_{sep+nonsep}
=S^{(2,metric)}_{6,1}/S^{(2,metric)}_{6,0}$) 
of the
34-dimensional hyperarea to the 35-dimensional volume for the seven metrics
under study, and the corresponding ratios for a 35-dimensional {\it Euclidean} 
ball having: (1) 
the same volume as for the metric;  and (2)  the same hyperarea}
\begin{ruledtabular}
\begin{tabular}{rrrrr}
metric & known ratio & sample ratio ($R^{metric}_{sep+nonsep}$) & isovolumetric ratio & isohyperarea ratio \\
\hline
Bures & 28.1205  & 28.1804   & 2.34553  & 2.40508   \\
KM & ---  & 0.0394299 & 1245.79     & 1536.34   \\
arith & ---  & 1.41531   & 38.858   &  43.2743  \\
WY & ---   & 2.08556  & 27.5655   &  30.39  \\
GKS & ---  & 8.07163 & 7.61987    & 8.08886  \\
geom & --- & $1.44625   \cdot 10^{-29}$ & $2.45463   \cdot 10^{29} $ &  $1.79638   \cdot 10^{30}$ \\
HS & 191.703  & 192.468  & 0.543466  & 0.533806 \\
\end{tabular}
\end{ruledtabular}
\end{table}
\subsection{Separability-probability ratios} \label{ratios}
\subsubsection{The $N=6$ qubit-qutrit case}
\begin{figure}
\includegraphics{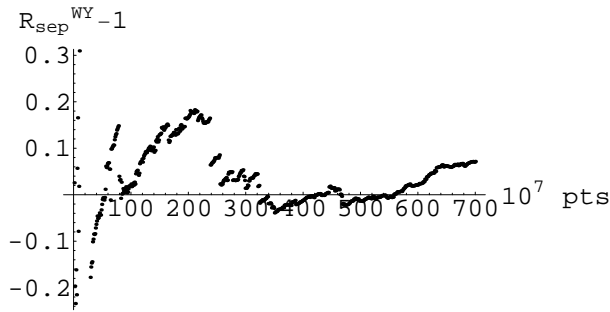}
\caption{\label{fig:graphratiosepWY}Deviations of the cumulative estimates of
$R^{sep}_{WY}$ from 1.}
\end{figure}

In Table~\ref{tab:isoperimetric2} we list for the seven metrics the 
estimated ratios, which we denote
$R^{metric}_{sep}$,  of
the hyperarea (consisting of only the rank-five but not the rank-six $6 \times 6$ density matrices constituting the boundary of the
{\it separable} 
density matrices \cite{shidu}) to the volume of all the separable states themselves.
\begin{table}
\caption{\label{tab:isoperimetric2}Sample estimates of the ratio ($R^{metric}_{sep}$) 
of the
34-dimensional hyperarea consisting only 
of rank-five $6 \times 6$ 
{\it separable} density 
matrices to the 35-dimensional separable volume for the seven metrics
under study. In the last column there are given the ratios of 
ratios ($\Omega^{metric}$)  of the middle (third)
column of Table~\ref{tab:isoperimetric} to these values}
\begin{ruledtabular}
\begin{tabular}{rrr}
metric & $R^{metric}_{sep}$ & $\Omega^{metric} \equiv 
R^{metric}_{sep+nonsep}/R^{metric}_{sep} 
=P_{6}^{[metric,6]}/P_{6}^{[metric,5]}$ \\
\hline
Bures & 14.501   & 1.94334    \\
KM & 0.0175377  & 2.24829  \\
arith &  0.692291  & 2.04439   \\
WY &  1.07164   & 1.94613   \\
GKS  & 4.14819  & 1.94582\\
geom &  $1.28502    \cdot 10^{-25}$ & 0.000112547   \\
HS & 94.9063   & 2.0279  \\
\end{tabular}
\end{ruledtabular}
\end{table}
We see that $R^{sep}_{WY}$ is quite close to 1. 
(The Wigner-Yanase metric is one of constant curvature \cite{gi}.)
In Fig.~\ref{fig:graphratiosepWY} we show the deviations of the cumulative 
estimates of $R^{sep}_{WY}$ from 1.
In the last column of Table~\ref{tab:isoperimetric2} 
there are given the ratios of 
ratios $\Omega^{metric} \equiv 
R^{metric}_{sep+nonsep}/R^{metric}_{sep}$. (The 
exceptional [geometric average] case might possibly simply be dismissed for serious consideration 
on the basis of numerical instabilities, with the associated volumes 
for this metric appearing to be actually
 {\it infinite} in nature. Also, as we will see below, $\Omega^{KM}$ is subject to particular severe jumps, perhaps decreasing the reliability of the
 estimates. The other five are rather close to 2 --- but it is also somewhat intriguing that three of the 
estimated monotone metric 
ratios are quite close to one another ($\approx 1.945$), 
 and therefore perhaps a common value
{\it unequal} to 2.)
This ratio-of-ratios can be easily be rewritten --- as explicated in the 
Introduction --- to take the form of a
ratio of separability
probabilities. That is, $\Omega^{metric}$  is equivalently the ratio
of the probability of separability ($P_{6}^{[(metric,6]}$) 
for all qubit-qutrit states to the 
conditional 
probability of separability ($P_{6}^{[metric,5]}$) for those states on the (rank-five) boundary of the
35-dimensional convex set. 

An interesting conjecture now would be that this ratio ($\Omega^{metric}$) 
is equal to the integral value 2,
{\it independently} of the (monotone or HS) metric used to measure the volumes and hyperareas.
If, in fact, valid, then there is presumably a {\it topological}
explanation \cite{witten}  for this. (We were able to quite readily reject
the speculation that this
phenomenon might be in some way an {\it artifact} of our particular
experimental design, in that we employ, as previously discussed, 
only for simple computational 
convenience, a 34-dimensional subsequence of
the 35-dimensional Tezuka-Faure sequence --- rather than an {\it ab initio}
independent 34-dimensional Tezuka-Faure sequence for the calculation of the
hyperareas.)

We must observe, however, that all the seven  
metrics specifically studied here induce
the {\it same} (Haar) measure over 30 of the 35 variables --- that is, the
30 Euler angles parameterizing the unitary matrices 
\cite{sudarshan,toddecg} --- but not over the five 
independent eigenvalues of the $6 \times
6$ density matrix. Therefore, it is certainly valid to point out 
that we 
have not considered {\it all} types of 
possible metrics over the 35-dimensional
space, but have restricted attention only to certain of those that are 
{\it not} inconsistent with quantum mechanical principles.
(S. Stolz has pointed out, in a personal communication  that, in general,
 one could  modify a metric in the interior away from the boundary {\it and}
outside the separable states,
without affecting the metric on the separable states, 
thus changing $R^{metric}_{sep+nonsep}$
without changing $R^{metric}_{sep}$, but obviously then also
altering the ratio of ratios ({\it proportio proportionum} \cite{oresme}) 
 $\Omega^{metric}$.
But presumably such a modification 
would lead, in our context, to the volume element of the 
so-modified metric {\it not} respecting Haar measure [cf. \cite[App. A]{hayden}].)

The {\it topology} of the $(N^2-1)$-dimensional convex set
of $N \times N$ density matrices has been laid out by \.Zyczkowski
and S{\l}omczynski \cite[sec. 2.1]{slom}.
The topological structure is expressible as
\begin{equation}
[U(N)/T^N] \times G_{N},
\end{equation}
where the group of unitary matrices of size $N$ is denoted by $U(N)$ and the
unit circle (one-dimensional torus $\approx U(1)$) by $T$, while $G_{N}$ 
represents an $(N-1)$-dimensional {\it asymmetric}
simplex. It would appear, however, that the set of separable states lacks
such a product topological structure (thus, rendering integrations over the 
set --- and hence the computation of 
corresponding volumes --- quite problematical).

\begin{figure}
\includegraphics{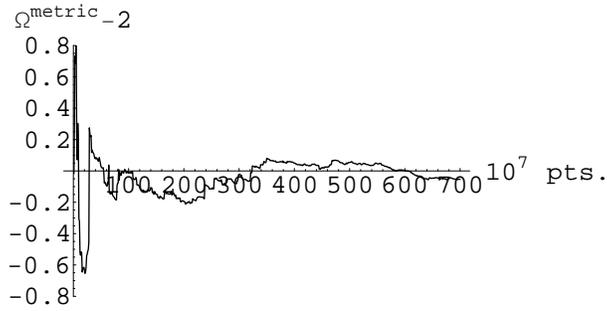}
\caption{\label{fig:ratioofratiosGKS}Deviations 
from the conjectured value of 2 of the cumulative
estimates of $R_{GKS}$, the  ratio of
hyperarea-to-volume
ratios for the Grosse-Krattenthaler-Slater (GKS or ``quasi-Bures'') 
  monotone metric}
\end{figure}

\begin{figure}
\includegraphics{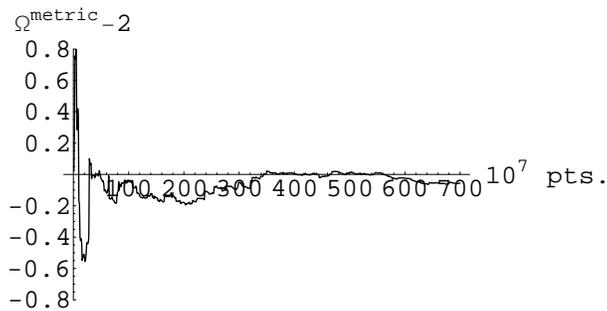}
\caption{\label{fig:ratioofratiosBures}Deviations from the conjectured value of 2 of the cumulative 
estimates of $R_{Bures}$, the  ratio of
hyperarea-to-volume
ratios for the Bures  monotone metric}
\end{figure}

\begin{figure}
\includegraphics{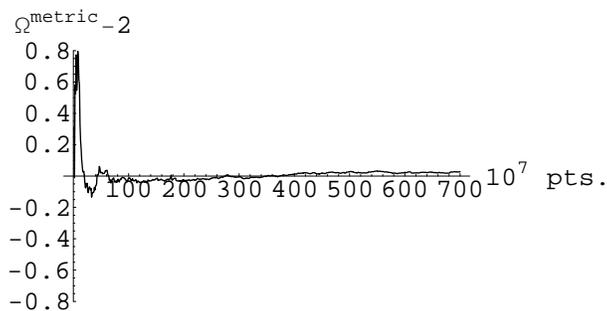}
\caption{\label{fig:ratioofratiosHS}Deviations from the conjectured value of 
2 of the cumulative 
estimates of $R_{HS}$, the ratio of
hyperarea-to-volume
ratios for the Hilbert-Schmidt metric. This plot is particularly flat in character}
\end{figure}

\begin{figure}
\includegraphics{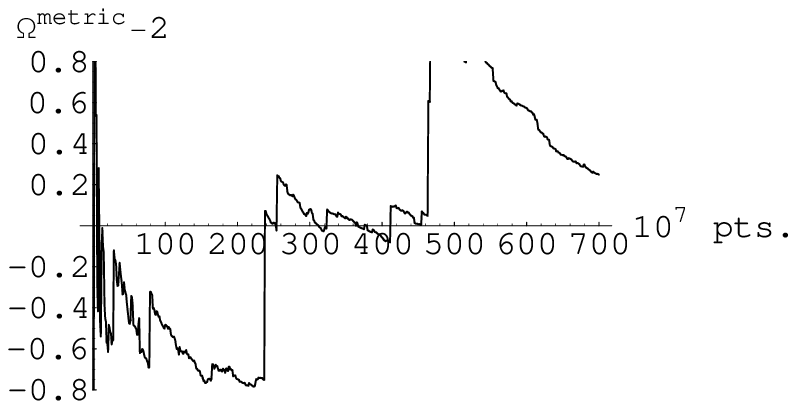}
\caption{\label{fig:ratioofratiosKM}Deviations from the conjectured value of 
2 of the cumulative estimates of $R_{KM}$, the ratio of 
hyperarea-to-volume
ratios for the KM (Kubo-Mori)  monotone metric}
\end{figure}

In Fig.~\ref{fig:ratioofratiosGKS} is plotted the deviations from the conjectured integral value of 2 of the cumulative estimates of
the ratio ($\Omega^{GKS}$) --- given in Table~\ref{tab:isoperimetric2} --- of 
the two hyperarea-to-volume ratios for the GKS 
monotone metric, 
the numerator ($R^{sep+nonsep}_{GKS}$) 
of $\Omega^{GKS}$ being based on 
the entirety of qubit-qutrit states, and the
denominator ($R^{sep}_{GKS}$) being based on the boundary qubit-qutrit 
states only. (All the succeeding plots of deviations from the conjectured integral value of 2 will be drawn to the {\it same} scale.)
In Figures~\ref{fig:ratioofratiosBures}, \ref{fig:ratioofratiosHS} 
and \ref{fig:ratioofratiosKM},
we show the corresponding plots based on the Bures, Hilbert-Schmidt
and 
 Kubo-Mori metrics, respectively.
We note that the cumulative estimates in this last plot were relatively close
to 2, before a sudden spike in the curve drove it upward.
The values for the (quite unrelated) Bures and HS-metrics are rather 
 close to 2, 
which is the main factor in our advancing the conjecture in question.

It would, of course, be of interest to study comparable ratios involving 
$6 \times 6$ density matrices of generic 
rank less than 5. We did not originally 
incorporate these into our MATHEMATICA 
Tezuka-Faure 
calculations (in particular, since we did not anticipate the apparent metric-independent phenomenon, we have 
observed  here). In sec.~\ref{newestone} below, we have, however, subsequently pursued such analyses.
\subsubsection{The $N=4$ qubit-qubit case} \label{N4qubitqubit}

\begin{figure}
\includegraphics{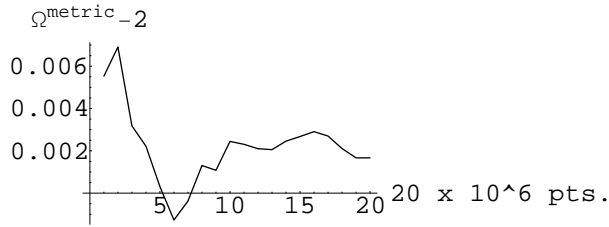}
\caption{\label{fig:graphRank4HS}Deviations 
from the possible true value of 
2 of the cumulative estimates of the ratio of the rank-4 
Hilbert-Schmidt  separability
probability to the rank-3 separability probability in the $N=4$ qubit-qubit
case. Note the greatly reduced scale of the $y$-axis.}
\end{figure}

We adapted our MATHEMATICA routine used so far
for the scenario $N=6$, so that it
would yield analogous results for $N=4$. Based on 400 million points of
a new independent 15-dimensional 
Tezuka-Faure sequence, we obtained the results reported
in Table~\ref{tab:reducedcase}. (We now use the lower-case counterparts of
the symbols $R$ and $\Omega$ to differentiate the $N=4$ case from the
$N=6$ one.)
\begin{table}
\caption{\label{tab:reducedcase}Counterparts for the qubit-{\it qubit} 
case $N=4$ of the
ratios of separability probabilities, based on 400 million points of a Tezuka-Faure sequence}
\begin{ruledtabular}
\begin{tabular}{rrrr}
metric & $r^{metric}_{sep+nonsep}$ & $r^{metric}_{sep}$ & $\omega^{metric}$ \\
\hline
Bures & 12.1563 & 6.58246  & 1.84676 \\
KM    & 0.506688 & 0.348945    & 1.45206  \\
arith & 2.19634 & 1.2269  & 1.79015  \\
WY  & 2.93791  & 1.73028  & 1.69794   \\
GKS & 6.03321 & 3.3661 & 1.79234 \\
geom & $4.02853   \cdot 10^{-16}$ & $7.1263  \cdot 10^{-16} $& 0.565304 \\
HS  &  51.9626   & 25.9596   & 2.00167 \\
\end{tabular}
\end{ruledtabular}
\end{table}
Here, once more, the ratios-of-ratios ($\omega^{metric}$) 
tend to show rather similar values,
with the two exceptional cases again being the geometric average metric
(which we suspect ---like the maximal (Yuen-Lax) monotone metric, from 
which it is partially formed --- simply 
gives infinite volumes and hyperareas) and
the somewhat unstable KM monotone metric (which now gives an atypically
{\it low} value!). We were somewhat surprised that
the Hilbert-Schmidt metric again gives, as for $N=6$, 
 a value quite close to 2. 
In Fig.~\ref{fig:graphRank4HS} we show (on a comparatively very fine scale) 
the deviations from 2 of the
cumulative estimates of the ratio of the Hilbert-Schmidt separability probability for the rank-4 states to that for the rank-3 states.

However, it now seems fairly certain 
that if there is a true common value for $\omega^{metric}$ across the metrics, then it is not an integral one (and thus possibly not a {\it topological} explanation).
The theoretical values predicted by (\ref{gammaBures}) and
(\ref{gammaHS}) for $r^{Bures}_{sep+nonsep}$
and $r^{HS}_{sep+nonsep}$ are $16384/(429 \pi) \approx 12.1566$ and
$30 \sqrt{3} \approx 51.9615$, respectively.
Also, consulting Table 6 of our earlier study 
\cite{slatersilver}, we find that 
using the conjectured and
known values for the qubit-{\it qubit} case ($N=4$) presented there 
gives us $\omega^{Bures} 
= 8192/(1419 \pi) \approx 1.83763$ and a somewhat similar  numerical 
value 
$\omega^{arith} = 408260608/73153125 \pi
\approx 1.77646$.

Let us also indicate in passing that this new independent Tezuka-Faure sequence yields estimates that are quite close to previously known and conjectured values. For example, the ratios of the estimates of $S^{(2,Bures)}_{4,0}$
and $S^{(2,Bures)}_{4,1}$ to their respective known values are 1.0001 and
.9999. Further, the ratios of the estimates of $\Sigma^{(2,Bures)}_{4,0}$
and $\Sigma^{(2,Bures)}_{4,1}$ (our estimate being 
0.138119) to their respective 
conjectured  values are 1.0001 and .99999.

Let us take this opportunity to note that our analyses here indicate that the conjectures given in Table 6 of \cite{slatersilver} for the 14-dimensional hyperareas --- denoted $B^s$ and $B^{s+n}$ there ---  appear 
to have been too large by a factor of 8.
\subsubsection{The $N=9$ rank-9 and-8 cases}

\begin{figure}
\includegraphics{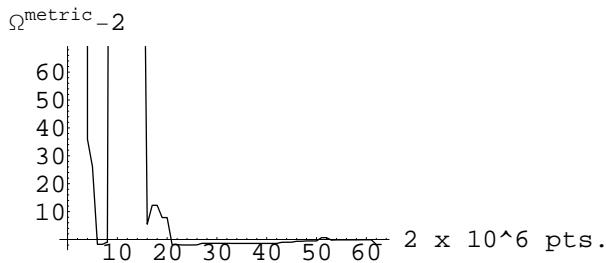}
\caption{\label{fig:graphRank9}Deviations from the value 
2 of the cumulative estimates of the ratio of the Hilbert-Schmidt 
probability of having a positive partial transpose for the  $9 \times 9$ density matrices of rank 9 to the probability in the rank-8 case}
\end{figure}

K. \.Zyczkowski has indicated to us that he has an argument, if not yet fully rigorous, to the effect that the ratio of the probability of rank-$N$ states having positive partial transposes to the probability of such rank-$(N-1)$ states should be 2 {\it independently} of $N$.
Some early analyses of our --- based on a 
so-far relatively short Tezuka-Faure sequence of 126 million points in the extraordinarily high (80)-dimensional 
hypercube --- gave us a 
Hilbert-Schmidt rank-9/rank-8 
probability 
ratio of 1.89125. (The analogous 
ratios for the monotone metrics were largely on the order
of 0.15. In these same analyses we also --- for our first time --- implemented, as well, the computable cross-norm criterion for separability
\cite{rudolph}, 
and found that {\it many} more density matrices could not be ruled out 
as possibly separable than with the [apparently much more discriminating] 
positive 
partial transposition criterion. The Hilbert-Schmidt 
probability ratio  based on the cross-norm criterion was 0.223149.). 
In Fig.~\ref{fig:graphRank9} we show the deviations from 2 of the 
cumulative estimates of the Hilbert-Schmidt rank-9/rank-8 ratio based on the positivity of the partial transpose.
However, this plot seems very unstable, so we must be quite cautious 
(pending a much more extended analysis) in
its interpretation.

\subsubsection{The $N=6$ rank-4 and rank-3 cases} \label{newestone}
The principal analyses above have been concerned with the full rank 
(rank-6) and rank-5 $6 \times 6$ density matrices. 
We adapted our MATHEMATICA procedure so that it would analyze the rank-4 and rank-3 cases, in a similar fashion. Now, we are dealing with 31-dimensional and 26-dimensional scenarios, {\it vis \`a vis} 
the original 35- and 34-dimensional ones.

In a preliminary run, based on 35 million points of corresponding Tezuka-Faure sequences, not a single rank-3 separable $6 \times 6$ density matrix was
generated. (The general results of Lockhart \cite{lockhart} --- based on
Sard's Theorem --- tells us 
that 
the measures of rank-2 and rank-1 $6 \times 6$ separable density matrices must be
zero, but not rank-3, as it appears we have observed [or near to it].) 
At that stage, we decided to concentrate further in our calculations
on the rank-4 case alone.

\begin{figure}
\includegraphics{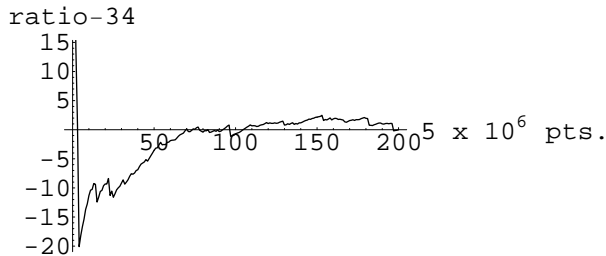}
\caption{\label{fig:graphDev34}Deviations 
from 34 of the cumulative estimates of the 
ratio of the rank-6 separability probability
for the HS-metric to the rank-4 separability probability}
\end{figure}

In Table~\ref{tab:N6rank4} we report results based on one  billion points of
a (new/independent) 31-dimensional Tezuka-Faure sequence, coupled with our estimates obtained
on the basis of our principal analysis, using the before-mentioned seven billion points.
\begin{table}
\caption{\label{tab:N6rank4}Estimated ratios of both 
rank-6 and rank-5 qubit-qutrit separability probabilities to rank-4 separability probabilities}
\begin{ruledtabular}
\begin{tabular}{rrr}
metric & rank-6/rank-4 ratio & rank-5/rank-4 ratio \\
\hline
Bures & 20.9605 & 10.7858 \\
KM & 12.2764  & 5.4603 \\
arith & 17.4245  & 8.52308 \\
WY & 15.5015  & 7.96527 \\
GKS & 18.3778  & 9.44474 \\
geom & $1.30244 \cdot 10^{-7}$ &  0.00115724\\
HS & 33.9982 & 16.7652  \\
\end{tabular}
\end{ruledtabular}
\end{table}
We note that for the Hilbert-Schmidt metric, 33.9982 (2 $\times$ 16.9991)
is quite close to integral. 
In Fig.~\ref{fig:graphDev34} we show the cumulative estimates of the ratio from the value 34.
(Of course, if the ratio of the rank-6 HS separability probability to the rank-5 HS separability probability is exactly, in theory, equal to 2, and the rank-6/rank-4 separability probability exactly 34, then
the rank-5/rank-4 ratio should be 17. Since it is 
 based on greater numbers of sampled separable density matrices, 
we suspect the sample  estimate of the rank-6/rank-4 HS separability
probability may perhaps be superior to the 
[less closely integral in value --- that is, 16.7652] rank-5/rank-4 estimate.)

Though the convergence to the predicted Hilbert-Schmidt volume was quite
good (99.9654\% of that given by the \.Zyczkowski-Sommers  formula \cite[eq. (5.3)]{hilb2}, 
for $N=6,n=2$),
we were initially 
rather disappointed/surprised that the predicted value of the Bures volume was inaccurate by some 25\%. This indicated to us that either 
the numerics were much more difficult for the Bures computation, 
or there was a possible error in our programming
(which we were unable to locate) or even the 
possibility that something was incorrect with the {\it specific}
Sommers-\.Zyczkowski formula \cite[eq. (4.19)]{hansrecent} we were using,
\begin{equation} \label{HZ2}
S^{(2,Bures)}_{N,n}= 2^{-d_{n}} \frac{\pi^{(d_{n}+1)/2}}{\Gamma((d_n+1)/2)}
\binom{N+n-1}{n}.
\end{equation}
This last possibility, in fact, proved to be the case, as we found that
their formula 
(4.19) did not agree (for cases other than $n=0,1,N-1$) 
with the 
more general formula (5.15) --- reproduced above as (\ref{HZ1}) --- and 
that using the correct formulation (5.15) (which we found also agrees with
(4.18) of \cite{hansrecent})
with $\beta=2,n=2,N=6$, 
our numerical deviation was reduced from 25\% to a more acceptable/less surprising  .1\%. A rectified version of their formula (4.19),
\begin{equation}
S^{(2,Bures)}_{N,n}=2^{-d_n}\frac{\pi^{(d_n+1)/2}}
{\Gamma((d_n+1)/2)}
\Pi_{j=0}^{N-n-1} \frac{j! (j+2n)!}{[(j+n)!]^2} ,
\end{equation}
has since been posted
by Sommers and \.Zyczkowski (quant-ph/0304041 v3) after this 
matter was brought to their attention.

\subsection{Well-fitting 
formulas for the Bures and Hilbert-Schmidt separable volumes and hyperareas} \label{wellfitting}
\subsubsection{The Bures case}
Proceeding under the assumption of the validity of our
conjecture above (regarding the integral value of 2 for 
$\Omega^{metric}$), 
computational experimentation indicates that the Tezuka-Faure quasi-Monte Carlo separable Bures 
results 
can be quite well fitted by taking
for the {\it separable} Bures 35-dimensional volume
\begin{equation} \label{sepconj60}
\Sigma^{(2,Bures)}_{6,0}= \frac{3 c_{Bures}}{2^{77}} \approx 1.03447 
\cdot 10^{-19},
\end{equation}
and for the {\it separable} (rank-five) 34-dimensional 
hyperarea 
\begin{equation} \label{sepconj61}
\Sigma^{(2,Bures)}_{6,1}= (2^{43} \cdot 3 \cdot 5 c_{Bures})^{-1} \approx 1.45449 \cdot 10^{-18},
\end{equation}
where by $\Sigma$ we denote volumes and hyperareas of {\it separable} states 
and 
\begin{equation}
c_{Bures} =  \sqrt{8642986 \pi} = \sqrt{\pi \cdot 2 \cdot 11 \cdot 19 \cdot 23 
\cdot 29 \cdot 31} \approx 5210.83.
\end{equation}
The pooled {\it sample} estimates, $\tilde{\Sigma}^{(2,Bures)}_{6,0}$ and 
$\tilde{\Sigma}^{(2,Bures)}_{6,1}$, as indicated in the caption to 
Table~\ref{tab:table7}, are $1.0142 \cdot 10^{-19}$ and 
$1.4707 \cdot 10^{-18}$.

Then, we would have the Bures probability of separability of the 
(generically rank-six) qubit-qutrit
states as
\begin{equation} \label{Buresprob}
P_{6}^{[Bures,rank-6]} = 
\frac{3^7 \cdot 5^3 \cdot 7^2 \cdot 11 \cdot 13 \cdot 17 c_{Bures}}{2^{27} \pi^{18}} \approx 0.00142278,
\end{equation}
and the Bures probability of separability of the generically rank-five
qubit-qutrit states, exactly (by our integral conjecture) 
one-half of this.
(However, there appears to be no obvious way in which the formulas immediately
above extend the 
analogous ones in the qubit-qubit separable 
case \cite{slatersilver}, which
were hypothesized to involve the ``silver mean'', $\sigma_{Ag}=\sqrt{2}-1$.
 Thus, it does not seem readily 
possible to use the results here to, in any way, 
support our earlier conjectures.)

We have also devised another set of exact Bures formulas that fit our data
roughly as well as (\ref{sepconj60}) and (\ref{sepconj61}).
These are 
\begin{equation} \label{ssepconj60}
\Sigma^{(2,Bures)}_{6,0}=  \frac{3^2 \cdot 11 \cdot 19 \cdot 23 \cdot 29 \cdot 31 \pi}{2^{76} \cdot 5^6} \approx 1.03497 
\cdot 10^{-19}
\end{equation}
and for the {\it separable} (rank-five) 34-dimensional
hyperarea
\begin{equation} \label{ssepconj61}
\Sigma^{(2,Bures)}_{6,1}= (2^{43} \cdot 5^7)^{-1} \approx 1.45519 \cdot
10^{-18}.
\end{equation}
\subsubsection{The Hilbert-Schmidt case}
Additionally, we can achieve {\it excellent} fits to our 
Hilbert-Schmidt estimates
by taking for the {\it separable} (rank-six) 35-dimensional volume
\begin{equation} \label{Sepconj60}
\Sigma^{(2,HS)}_{6,0}= (2^{45} \cdot 3 \cdot 5^{13} \cdot 7 \sqrt{30})^{-1}
\approx 2.02423 \cdot 10^{-25}
\end{equation}
(the sample estimate being $2.05328 \cdot 10^{-25}$)
and for the {\it separable} (rank-five) 34-dimensional
hyperarea
\begin{equation} \label{Sepconj61}
\Sigma^{(2,HS)}_{6,1}= (2^{46} \cdot 3 \cdot 5^{12})^{-1} 
\approx 1.94026 \cdot 10^{-23}
\end{equation}
(the sample estimate being $1.94869    \cdot 10^{-23}$).
This gives us a Hilbert-Schmidt probability of separability of the
generically rank-six states of
\begin{equation}
P_{6}^{[HS,rank-6]}= \frac{3^{10} \cdot 7^4 \cdot 11^3 \cdot 13^2 \cdot 17^2 \cdot 19 \cdot 23 \cdot 29 \cdot 31 \sqrt{5}}{2^{37} \cdot 5^7 \pi^{15}} \approx 0.0263115,
\end{equation}
approximately 18.5 times the predicted Bures probability (\ref{Buresprob}).
(As we have noted, the Bures separability probability appears to be the
{\it greatest} among the monotone metrics.) An {\it upper} bound of
$0.166083 \approx (0.95)^{35}$ on $P_{6}^{[HS,rank-6]}$ is given in Appendix G of \cite{szarek}.

Our simple 
excellent Hilbert-Schmidt fits here led us to investigate whether the same could be achieved in the qubit-qubit ($N=4$) case, using the same 
400 million point Tezuka-Faure sequence employed in sec.~\ref{N4qubitqubit}.
This, in fact, seemed definitely doable, by taking
\begin{equation}
\Sigma^{(2,HS)}_{4,0} = (3^3 5^7 \sqrt{3})^{-1} \approx 2.73707 \cdot 10^{-7},
\end{equation}
(the sample estimate being $2.73928 \cdot 10^{-7}$)
and
\begin{equation}
\Sigma^{(2,HS)}_{4,1}= (3^2 5^6)^{-1} \approx 7.11111 \cdot 10^{-6}
\end{equation}
(the sample estimate being $7.11109 \cdot 10^{-6}$).
The {\it exact} 
Hilbert-Schmidt probability of separability of the generically rank-4 
qubit-qubit states would then be
\begin{equation}
P_{4}^{[HS,rank-4]}= \frac{2^2 \cdot 3 \cdot 7^2 \cdot 11 \cdot 13 \sqrt{3}}{5^3 \pi^6} \approx
0.242379.
\end{equation}
(A {\it lower} bound for $\Sigma_{4,0}^{(2,HS)}$ of 
${256 \pi^7/29805593211675} \approx 2.65834 \cdot 10^{-8}$ --- that is, the volume of a 15-dimensional ball of radius $\frac{1}{3}$ --- appears obtainable
from the results of \cite{sepsize3}, although it is not fully clear 
to this reader whether 
the argument there applies to the {\it two}-qubit case [$m=2$], since the exponent $\frac{m}{2}-1$ appears.)
Of course, one would now like to try to extend these Hilbert-Schmidt results
to cases $N>6$.
Also, let us propose the formula
\begin{equation}
\Sigma^{(2,HS)}_{6,2} = \frac{7 \cdot 11}{2^{41} 5^{11} \sqrt{5} \pi} \approx 
1.02084 \cdot 10^{-19},
\end{equation}
(the sample estimate [sec.~\ref{newestone}] being $1.04058 \cdot 10^{-19}$).

\section{Discussion}

In the main numerical analysis of 
this study (secs.~\ref{secMonotone} and \ref{secHS}), 
we have directly estimated 28 quantities of interest --- seven
total volumes of the 35-dimensional space of qubit-qutrit states, seven
34-dimensional hyperareas of the boundary of those states, and the same
quantities when restricted to the separable qubit-qutrit states.
Of these 28 quantities, four
(that is, $S^{(2,Bures)}_{6,0}, S^{(2,Bures)}_{6,1}, S^{(2,HS)}_{6,0}$ 
and $S^{(2,HS)}_{6,1}$) were 
precisely 
known from previous analyses of Sommers
and \.Zyczkowski \cite{hans1,hilb2}. 
It is interesting to observe that the Tezuka-Faure
quasi-Monte Carlo numerical integration procedure has, in all four  of these
cases, as shown in the corresponding table and figures
(Table~\ref{tab:table1} and Figs.~\ref{fig:graphBuresTotVol}, \ref{fig:graphBuresTotArea}, \ref{fig:graphHSTotVol}, \ref{fig:graphHSTotArea}), 
slightly but consistently 
{\it under}estimated the known values --- more so, it seems, with
the 35-dimensional volumes, as opposed to the 34-dimensional 
hyperareas. (So, in statistical terminology, we 
appear to have {\it biased} estimators. The very same form of bias --- in terms of the Bures metric --- was observed
in the precursor analysis \cite{slaterprecursor} to this one, based on
an independent, shorter Tezuka-Faure sequence. {\it Randomizing} deterministic algorithms --- such as the Tezuka-Faure --- can remove such bias \cite{hong}.) 
This suggests that we
might possibly 
improve the accuracy of the estimates of the 24 unknown quantities
by scaling them in accordance with the magnitude of 
known underestimation.
Also, we have in our several tables only reported the results at the 
(seven-billion point)
end of the Tezuka-Faure procedure. We might also report results at intermediate
stages at which the estimates of the 4 known quantities are closest to their
true values, since estimates of the 24 unknown quantities might arguably
also be most accurate at those stages.

Of course, as we have done, 
taking the {\it ratios} of estimates of the volumes/hyperareas
 of separable states to the estimates of the volumes/hyperareas of
separable plus nonseparable states, one, in turn,  obtains estimates of
the probabilities of separability \cite{ZHSL} 
for the various monotone metrics studied. ({\it Scaling} the estimated volumes
and hyperareas by the corresponding estimates for the  
Bures metric, as we have done in certain of the
tables above for numerical convenience and possible insightfulness,
would be inappropriate in such a process.) 
Among the metrics studied, 
the Hilbert-Schmidt metric gives the {\it largest} 
qubit-qutrit probability of separability 
($\approx 0.0268283$), while the Bures metric --- the {\it minimal} monotone one --- gives 
the (considerably smaller) 
largest separability probability ($\approx 0.00139925$) 
among the monotone metrics studied
(and presumably among all monotone metrics).  The (Yuen-Lax) {\it maximal}
monotone metric appears to give a null separability probability.

In \cite{qubitqutrit}, we had 
attempted a somewhat similar quasi-Monte Carlo 
qubit-qutrit 
analysis (but restricted simply 
to the Bures metric) to that reported above, but based on many fewer points (70 million {\it vs.} the 7 billion so far 
used here) of a 
(Halton) sequence.
At this stage, having made use of considerably increased computer power (and streamlined MATHEMATICA programming --- in particular 
employing the Compile command, which enables the program to 
proceed under the condition
that certain variables will enter a calculation only as 
machine numbers, and
not as lists, algebraic objects or any other kind of expression),
we must regard this earlier study as entirely superseded by the one here.
(Our pooled estimate of the Bures volume of the separable qubit-qutrit
systems here  [Table VII] 
is $1.0142  \cdot 10^{-19}$, while in \cite{qubitqutrit}, following our earlier work for $N=4$ \cite{firstqubitqubit}, 
we formulated a conjecture (\cite[eq. (5)]{qubitqutrit}) --- in which we can now have but 
very little
confidence --- that would give [converting from the SD metric to the 
Bures] a value of
 $2^{-35} \cdot (2.19053 \cdot 10^{-9}) \approx  6.37528 \cdot 10^{-20}$.)
We also anticipate revisiting --- as in sec.~\ref{N4qubitqubit} --- the 
$N=4$ (qubit-qubit) case \cite{slatersilver}
with our newly accelerated programming methods, in a similarly systematic 
manner.

Perhaps, in the future, subject to research priorities, we will 
add to the 7 billion points of the Tezuka-Faure sequence employed above, and hope to report considerably more accurate results in 
the future (based on which, possibly,
 we can further appraise the hypotheses offered above as to the values of the various 
volumes and hyperareas). Also, we may seek to 
estimate the hyperarea of that part
of the boundary of the separable qubit-qutrit states consisting of
generically rank-six $6 \times 6$ density matrices \cite{slatersilver,shidu},
though this involves a much greater amount of 
computation per point. (This would entail first
finding the values, if any, 
 of the undetermined [35-{\it th}] parameter that would set
the determinants of the two forms of 
partial transpose equal to zero, and then --- using these values --- 
ascertaining whether or not
all the six eigenvalues of the resultant partial transposes
were nonnegative.)

In this study, we have utilized additional computer power
recently available to us,
together with an advanced quasi-Monte Carlo procedure
(scrambled Faure-Tezuka sequences 
\cite{tezuka,giray1} --- the use of which was recommended to us
by G. \"Okten, who  provided a corresponding MATHEMATICA
code). Faure and Tezuka were guided ``by the construction $C^{(i)} = A^{(i)}
P^{(i-1)}$ and by some possible extensions of the generator formal series in
the framework of Neiderreiter''. ($A^{(i)}$ is an arbitrary nonsingular lower
triangular [NLT] matrix, $P$ is the Pascal matrix \cite{call}
and $C^{(i)}$ is a generator matrix of a sequence $X$.)
Their idea was to multiply from the right by
nonsingular upper triangular (NUT) random matrices and get the new generator
matrices $C^{(i)}= P^{(i-1)} U^{(i)}$ for $(0,s)$-sequences
\cite{tezuka,giray1}.``Faure-Tezuka scrambling
scrambles the digits of $i$ before multiplying by the generator matrices \ldots\
 The effect of the Faure-Tezuka-scrambling can be thought of as reordering
the original sequence, rather than permuting its digits like the Owen
scrambling \ldots Scrambled sequences often have smaller
discrepancies than their nonscrambled counterparts. Moreover, random
scramblings facilitate error estimation'' \cite[p. 107]{hong}.

It would be interesting to conduct analogous investigations to those reported here ($N=6$) 
and in \cite{slatersilver} for the case $N=4$, using
quasi-random sequences {\it other}
 than Tezuka-Faure ones \cite{tezuka,giray1}, particularly those for which it is possible to do {\it statistical} testing on the results
(such as constructing confidence intervals) \cite{hong}. It is, of course, 
possible to conduct statistical testing using simple Monte Carlo methods, but
their convergence is much weaker than that of the quasi-Monte Carlo
procedures. Since we have been dealing with 
extraordinarily high-dimensional spaces, good
convergence has been a  dominant consideration in the selection of
numerical integration methodologies to employ.

``It is easier to estimate the error of Monte Carlo methods because one can
perform a number of replications and compute the variance. Clever 
randomizations of quasi-Monte Carlo methods combine higher accuracy with practical
error estimates'' \cite[p. 95]{hong}.
 G. \"Okten is presently developing a new MATHEMATICA version of scrambled
Faure-Tezuka sequences in which
there will be
 a random generating matrix for each dimension --- rather than
one for all the dimensions together --- which will then be
susceptible to {\it statistical} testing \cite{hong}.
 
At the strong urging of K. \.Zyczkowski, we disaggregated
the pooled results in the last
column of Table IX into the part based on partial transposition of
four three-by-three blocks and obtained
$\{1.966, 2.53505, 2.04826, 1.94679, 1.96481, 9.32089 \times 10^-7, 1.99954\}$
and into
the part based on nine two-by-two blocks and obtained
$\{1.91846, 1.91976, 2.04, 1.94539, 1.92476, 0.0001227, 2.05803\}$.
We bring the attention of the reader to the particular
closeness to 2 of the first (Hilbert-Schmidt) ratio.

\begin{acknowledgments}
I wish to express gratitude to the Kavli Institute for Theoretical
Physics (KITP)
for computational support in this research and to Giray \"Okten
for supplying the MATHEMATICA code for the Tezuka-Faure quasi-Monte Carlo
procedure and for numerous communications. Chris Herzog of the KITP 
kindly provided certain computer assistance, 
as well as comments on the course of the research.
S. Stolz remarked on our conjecture regarding the integral value of 2. Also,
K. \.Zyczkowski supplied many useful comments and insights.

\end{acknowledgments}

\bibliography{HS3}

\end{document}